\documentclass{iau}
\usepackage{graphicx,natbib}

\newcommand{\Msunyr}{M$_{\odot}$yr$^{-1}$}
\newcommand{\Mjup}{M$_{\mathrm {Jup}}$}
\newcommand{\Msun}{M$_{\odot}$}

\newcommand{\apj}{{\it ApJ}}
\newcommand{\apjl}{{\it ApJL}}
\newcommand{\aap}{{\it A\&A}}
\newcommand{\mnras}{{\it MNRAS}}
\newcommand{\pasp}{{\it PASP}}
\newcommand{\aj}{{\it AJ}}
\newcommand{\pasa}{{\it PASA}}
\newcommand{\nar}{{\it New Astron.~Revs.}}
\newcommand{\araa}{{\it ARAA}}
\newcommand{\icarus}{{\it Icarus}}

\title[Planet formation in evolving protoplanetary discs] 
{Planet formation in evolving protoplanetary discs}

\author[Alexander]   
{Richard Alexander}

\affiliation{Department of Physics \&
  Astronomy, University of Leicester, Leicester, LE1 7RH, UK \\email: {\tt richard.alexander@leicester.ac.uk}}

\pubyear{2013}
\volume{299}  
\pagerange{1--10}
\jname{Exploring the formation and evolution of planetary systems}
\editors{B.Matthews \& J.Graham, eds.}
\begin{document}

\maketitle

\begin{abstract}
I attempt to summarize our knowledge of planet formation in evolving protoplanetary discs.  I first review the physics of disc evolution and dispersal. For most of the disc lifetime evolution is driven by  accretion and photoevaporation, and I discuss how the interplay between these processes shapes protoplanetary discs.  I also discuss the observations that we use to test these models, and the major uncertainties that remain.  I will then move on to consider planet formation and migration in evolving discs, and discuss how observations of both discs and planets can be used to inform our understanding of protoplanetary disc evolution.
\end{abstract}

\firstsection 


\section{Introduction}
\noindent We now know that many, if not most, stars host planetary systems, but their origins remain a mystery.  Understanding the evolution of protoplanetary discs is a critical ingredient of any predictive theory of planet formation. Protoplanetary discs are the sites of planet formation, but planet formation is an inefficient process (only a small fraction of the initial disc mass ends up in planets) and consequently discs dominate the dynamics of forming planetary systems. Here I review our understanding of the physics of protoplanetary disc evolution and dispersal, and its consequences for the architecture of planetary systems. 


\subsection{Disc evolution}
The existence of gas-rich protoplanetary discs was first suggested as long ago as the work of Kant \& Laplace, but such discs were not directly observed until the late 1980s \citep[e.g.,][]{sb87}.  Since then observations of discs have progressed at a startling rate, and we now have a census of thousands of these objects.  Discs are observed across a wide range of wavelengths, using both gas and dust tracers, and this has allowed us to build up a detailed picture of their properties.  Observations of young clusters find that the fraction of stars with an infrared excess (which traces warm dust at $\sim$AU radii) declines from $\sim$100\% at ages $\lesssim$1Myr to $\lesssim$10\% at ages of 5--10Myr \citep[e.g.,][]{haisch01,mamajek09}.  A similar trend is seen in signatures of accretion on to the stellar surface \citep[e.g.,][]{fedele10}, which trace gas in the inner disc, and in (sub-)mm continuum emission, which traces cold dust at large ($\sim$100AU) radii \citep[e.g.,][]{aw05}. Although age estimates for young stars, which are typically derived from comparing observations to pre-main-sequence evolutionary tracks, are subject to significant uncertainties \citep{hillenbrand09,bell13}, from these data we can conclude that typical disc lifetimes are a few Myr, with order-of-magnitude scatter.

In addition to statistical studies of discs in clusters, these observations also allow us to determine global properties of individual discs.  By making standard assumptions about the dust opacity and dust-to-gas ratio we can estimate disc masses from the (sub-)mm luminosity: typical disc masses range from $M_{\mathrm d}$\,$\sim$\,0.1\Msun\ to $\lesssim$\,0.001\Msun\ \citep[e.g.,][]{aw05}.  We can also derive instantaneous accretion rates, by measuring the excess short-wavelength emission produced by the accretion shock on the stellar surface; if we know the stellar mass and radius, we can convert an observed accretion luminosity into a mass accretion rate \citep[e.g.,][]{calvet98}.  Accretion rates measured in this manner (for $\sim$\,solar-mass stars) range from $\dot{M}_{\mathrm {acc}}$\,$\gtrsim$\,$10^{-7}$\,\Msunyr\ to $\lesssim$\,$10^{-10}$\Msunyr\ \citep[e.g.,][]{gullbring98}.  We can therefore estimate an empirical accretion timescale $t_{\mathrm {acc}}$\,$\sim$\,$M_{\mathrm d}/\dot{M}_{\mathrm {acc}}$, and find that for most sources $t_{\mathrm {acc}}$ is of order Myr, again with large scatter \citep[e.g.,][]{jones12}.  The accretion time-scale is approximately the time it will take all of the observed disc mass to be accreted (at the observed accretion rate), so we expect protoplanetary discs to evolve substantially due to accretion during their  $\sim$Myr lifetimes.  This in turn implies that there is no such thing as ``typical conditions'' for planet formation; planets form in evolving protoplanetary discs, and the ubiquity of planets suggests that they form under a wide range of different conditions.

These observations show that protoplanetary discs evolve, at least in part due to disc accretion, but shed little light on the physical processes driving this evolution.  The decline in observed accretion rates with stellar age is broadly consistent with classical viscous accretion disc evolution \citep[e.g.,][]{lbp74,hartmann98}, but the origin of the required angular momentum transport remains unclear \citep[see, e.g., the review by][]{armitage11}. At early times, when the disc is still embedded in its parent envelope, disc evolution is probably dominated by gravitational instabilities \citep[e.g.,][]{lodato07}. Magnetically-driven jets and winds can also remove angular momentum and drive accretion in the disc, particularly at small radii, and may remain important throughout the disc's evolution \citep[e.g.,][]{ks11}. However, for most of the disc lifetime the dominant source of angular momentum transport is thought to be magnetohydrodynamic (MHD) turbulence, driven by the magnetorotational instability \citep[MRI;][]{bh91,balbus11}. It remains unclear how much of the disc is sufficiently ionized to sustain the MRI, and it seems likely that substantial ``dead zones'', with little or no turbulence, exist in the majority of discs \citep[e.g.,][]{gammie96,bai11}. Recent work also suggests that MRI-driven winds may play a role in disc evolution, potentially driving significant mass- and angular momentum-loss over a wide radial extent \citep[e.g.,][]{fromang13,bai13}.  Thus, although simple viscous disc models are sufficient to explain most observations of protoplanetary disc accretion, our ignorance of how discs accrete remains a major obstacle to a complete theory of planet formation.


\subsection{Disc dispersal}
Observations of young stars also allow us to infer a great deal about disc dispersal. Different disc tracers are strongly correlated, despite originating in very different regions of the disc, and tend to vanish together; few objects are seen with ``partial'' discs \citep[e.g.,][]{aw05}. This suggests that disc dispersal is almost simultaneous across a wide range of radii, from $\lesssim$\,0.1AU to $>$\,100AU.  Moreover, the relative lack of objects with properties between those of disc-bearing and disc-less stars suggests that dispersal is rapid: statistical estimates suggest that the dispersal time-scale is at least an order of magnitude shorter than the disc lifetime \citep[e.g.,][]{sp95,koepferl13}. Finally, non-detections of warm or hot gas around disc-less weak-lined T Tauri stars set very strict upper limits on the gas surface density (orders of magnitude below canonical protoplanetary disc values), implying that disc dispersal is very efficient \citep{pascucci06,ingleby11}. Disc evolution therefore has two distinct time-scales (the $\sim$Myr disc lifetime and the much shorter disc dispersal time), and the process(es) which drive final disc clearing must operate very efficiently over the entire radial extent of the disc.

These observational constraints rule out several obvious candidates as mechanisms for disc dispersal. Resolved observations of disc structure suggest that most of the disc mass resides at large radii \citep[$\gtrsim$\,50AU;][]{andrews09}, but the characteristic viscous time-scale is $\gtrsim$Myr at these radii.  Discs therefore live for no more than a few viscous time-scales in their outer regions, and consequently accretion cannot play a major role in outer disc clearing.  Similarly, winds or jets driven by the stellar magnetic field cannot be the dominant disc clearing mechanism at large radii, though MRI-driven winds may play a role. However, in most discs the primary driver of final disc dispersal is thought to be disc photoevaporation \citep[see, e.g.,][]{clarke11,rda_ppvi}. 


\section{Disc evolution models}
\subsection{Disc Photoevaporation}
\noindent Photoevaporation occurs when high-energy (UV and/or X-ray) radiation is incident on protoplanetary discs. This irradiation creates a hot ($\sim$\,$10^3$--$10^4$K) layer on the disc surface, and beyond a certain radius the heated gas is unbound. The result is a pressure-driven flow, which is referred to as a photoevaporative wind. The characteristic critical radius depends on the temperature of the hot gas, and is approximately 
\begin{equation}
R_{\mathrm c} \simeq \frac{0.2GM_*}{c_{\mathrm {hot}}^2} \simeq 1.8 \left(\frac{M_*}{1\mathrm M_{\odot}}\right) \left(\frac{T_{\mathrm {hot}}}{10^4\mathrm K}\right)^{-1} \mathrm {AU} \, ,
\end{equation}
where $M_*$ is the stellar mass, and $c_{\mathrm {hot}}$ and $T_{\mathrm {hot}}$ are the sound speed and temperature in the heated surface layer, respectively \citep{hollenbach94,liffman03,font04}.  The irradiation can be both ``internal'' (i.e., from the central star) or ``external'' (from nearby massive stars); external heating dominates in some cases (e.g., the propldys in the centre of the Orion Nebula Cluster; \citealt{johnstone98}), but for the majority of discs we expect photoevaporation to be dominated by irradiation from the central star.

The temperature of the surface layers depends primarily on the irradiation, and three regimes are  important for protoplanetary discs: ionizing, ``extreme'' UV (EUV; 13.6--100eV); non-ionizing far-UV (FUV; 6--13.6eV) and X-rays (0.1--10keV):
\begin{itemize}
\item[{\bf EUV:}] In the EUV case heating and cooling is dominated by ionization and recombination of atomic hydrogen, and for an optically thick disc ionizing photons from recombinations in the disc atmosphere dominate the heating at the ionization front \citep{hollenbach94}. The resulting disc atmosphere is close to isothermal at $T_{\mathrm {hot}}$\,$\simeq$\,$10^4$K (akin to an H\,{\sc ii} region on the disc surface), and most of the mass-loss originates at 1--2AU. As the flow is recombination-limited the integrated mass-loss rate $\dot{M}_{\mathrm w}$ scales as the square-root of the ionizing luminosity, and for fiducial parameters we find  $\dot{M}_{\mathrm w}$\,$\sim$\,$10^{-10}$\Msunyr\ \citep{hollenbach94,font04}.  However, if the inner disc is optically thin to ionizing photons (as expected during disc dispersal) then direct irradiation of the inner disc edge dominates. In this case most of the mass-loss comes from close to the inner edge, and $\dot{M}_{\mathrm w}$ increases by a factor of $\sim$10 \citep{alexander06a}.
\item[{\bf FUV:}] FUV irradiation is analogous to the well-studied problem of photodissociation regions (PDRs): photons are predominantly absorbed by either dust grains or polycyclic aromatic hydrocarbons (PAHs), and the gas is heated by a combination of collisions with grains and FUV-pumping of H$_2$.  The resulting gas temperatures vary substantially with both radius and height above the disc midplane, from $\sim$\,100K at large radii to $\gtrsim$\,1000K close to the star \citep{adams04,gh08,gh09}. The resulting mass-loss profile has a peak at 5--10AU, and also rises towards the disc outer edge ($\gtrsim$\,100AU); in some cases the mass-loss at large radii dominates. The integrated wind rate depends mainly on the FUV luminosity incident on the disc, and for fiducial parameters $\dot{M}_{\mathrm w}$\,$\sim$\,$10^{-8}$\Msunyr\ \citep{gh09}. 
\item[{\bf X-rays:}] Young solar-mass stars have long been known to be bright X-ray sources, with typical luminosities $L_{\mathrm X}$\,$\sim$\,$10^{30}$erg\,s$^{-1}$ \citep[e.g.,][]{feigelson07}. X-rays are primarily absorbed by K-shell ionization of heavy elements, and the resulting $\sim$keV photoelectrons collisionally ionize and/or heat the disc gas.  As in the FUV case, the disc atmosphere is not well characterised by a single value of $T_{\mathrm {hot}}$, but hydrodynamic models find that most of the resulting wind originates in the atomic layer, at temperatures of 3000--5000K \citep{owen10}, and consequently the mass-loss profile peaks at around 3AU.  The mass-loss rate scales close to linearly with $L_{\mathrm X}$, and for typical parameters we again find $\dot{M}_{\mathrm w}$\,$\sim$\,$10^{-8}$\Msunyr\ \citep{owen11,owen12}.
\end{itemize}


\begin{figure}[t]
\begin{center}
 \includegraphics[width=3.4in,angle=270]{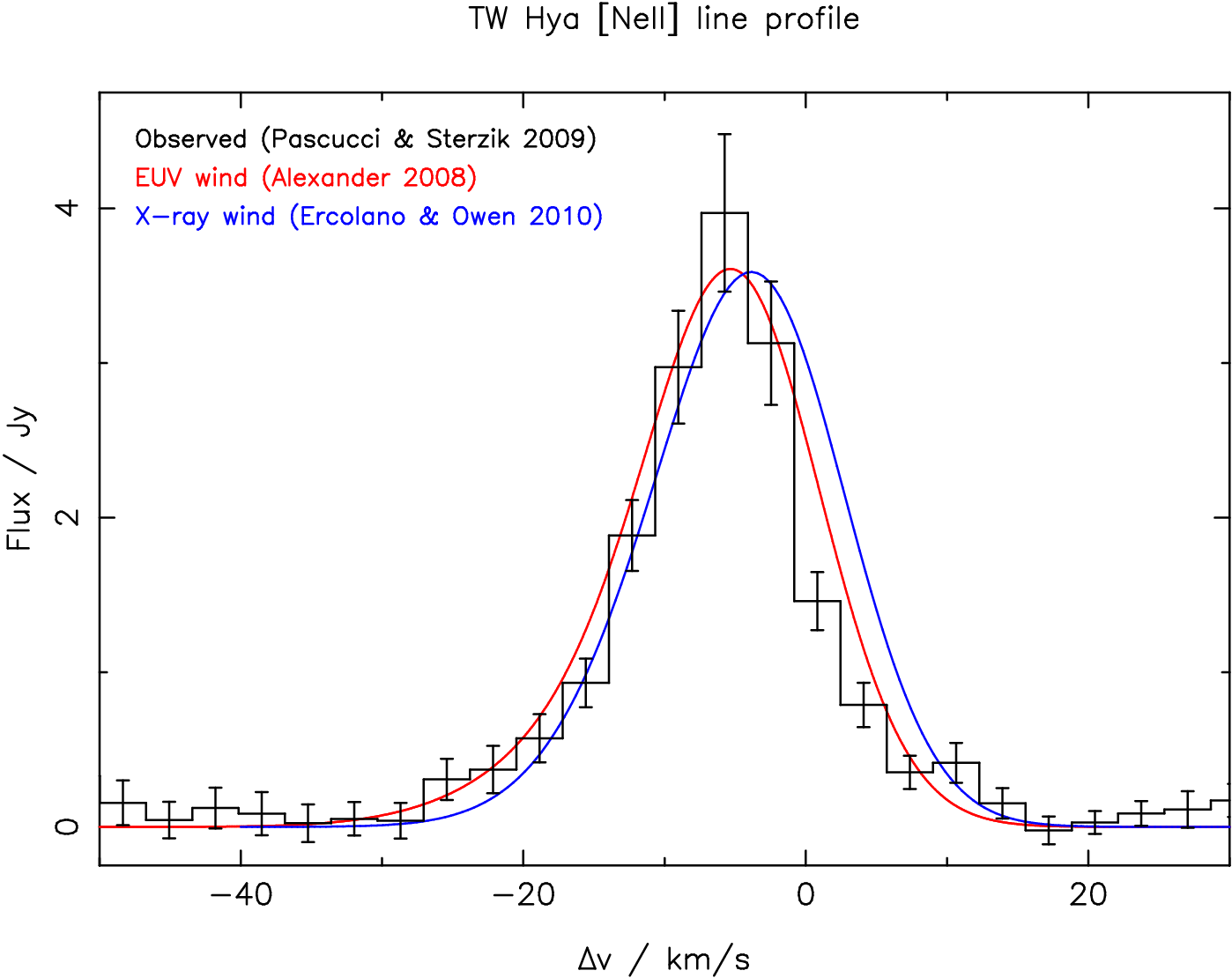} 
 \caption{TW Hya [Ne\,{\sc ii}] 12.81$\mu$m line profile: the black line shows the observed line profile from \citet{ps09}, while the red and blue curves show the predictions from photoevaporative wind models driven by stellar EUV \citep{alexander08b} or X-ray \citep{eo10} photons, respectively. Although the data do not yet distinguish between these models, the blue-shifted line represents a clear detection of a slow, ionized wind, and provides unambiguous evidence of on-going photoevaporation of the TW Hya disc.}
   \label{fig:neii}
\end{center}
\end{figure}


\vspace*{12pt}
\noindent These different ``flavours'' of photoevaporation models all predict mass-loss at rates that are large enough to play a significant role in disc evolution, but it is not clear which heating mechanism dominates in real discs.  Some theoretical issues remain, but the dominant uncertainty in these models is the irradiating flux which reaches the disc surface (which is an input parameter in the models).  Recent work has attempted to resolve this question observationally, by looking for emission line diagnostics which trace the wind structure.  Photoevaporative winds are characterised by relatively low-density gas which is at least partially ionized, and are therefore expected to produce copious forbidden line emission \citep{hg09}.  Hydrodynamic models show that the characteristic signature of such a wind is a small blue-shift in the emission from face-on discs \citep{font04}, and the [Ne\,{\sc ii}] 12.81$\mu$m line is arguably the most promising such diagnostic \citep{alexander08b}.  In this case the line originates close to the base of the wind, and models predict small blue-shifts ($\Delta v$\,$\sim$\,5--10km\,s$^{-1}$) for both EUV \& X-ray heated winds \citep{alexander08b,eo10}.  This blue-shift was first detected in the TW Hya disc \citep[][see Fig.\,\ref{fig:neii}]{ps09}, and has subsequently been seen in around a dozen other sources \citep{pascucci11,sacco12,baldovin12}.  The observed line profiles do not yet distinguish between the different photoevaporation models (see Fig.\,\ref{fig:neii}), but these observations represent an unambiguous detection of a slow wind which is at least partially ionized.  When combined with line profiles from other species (such as [O\,{\sc i}], e.g., \citealt{rigliaco13}), or other wind diagnostics such as free-free emission \citep{pascucci12}, these observations offer the prospect of measuring photoevaporation rates empirically.


\subsection{Secular evolution}
\begin{figure}[t]
\begin{center}
 \includegraphics[width=3.4in,angle=270]{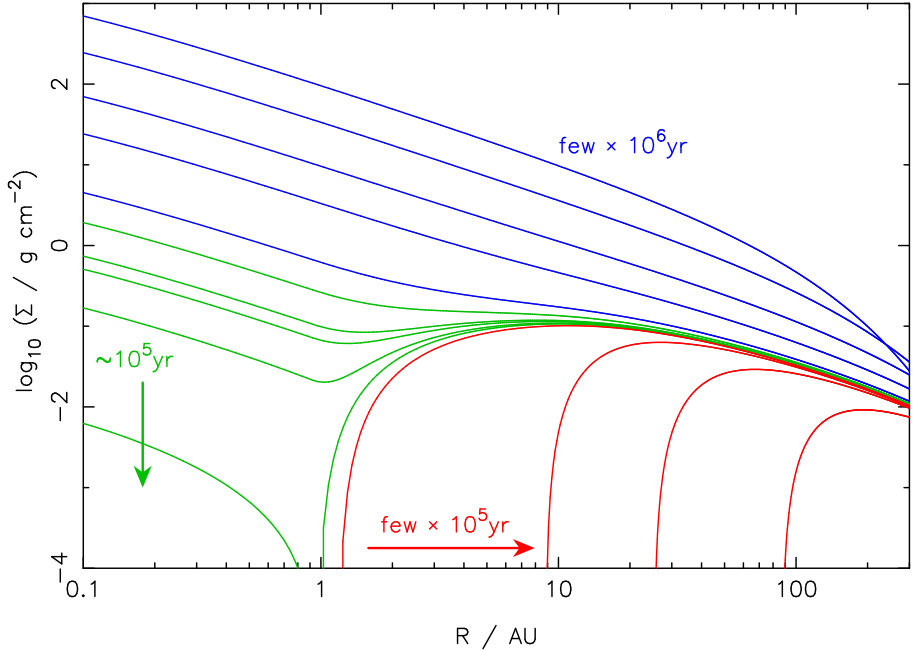} 
 \caption{Surface density evolution of a characteristic viscous/photoevaporation disc model, colour-coded to denote the three stages of the disc's evolution.  For most of the disc's lifetime (few Myr, blue curves) photoevaporation is negligible and the disc follows the behaviour of a classical viscous accretion disc. However, once the disc accretion rate drops sufficiently photoevaporation is able to open a gap in the disc at $\simeq$\,$R_{\mathrm c}$, after which the inner disc rapidly accretes on to the star ($\sim$\,$10^5$yr, green curves).  Once the inner disc becomes optically thin, photoevaporation clears the disc from inside out (few\,$\times$\,$10^5$yr, red curves).  [Figure adapted from \citet{alexander08a}, using the median disc model of \citet{aa09}.]}
   \label{fig:surf_den}
\end{center}
\end{figure}


\begin{figure}[t]
\begin{center}
 \includegraphics[width=3.4in,angle=270]{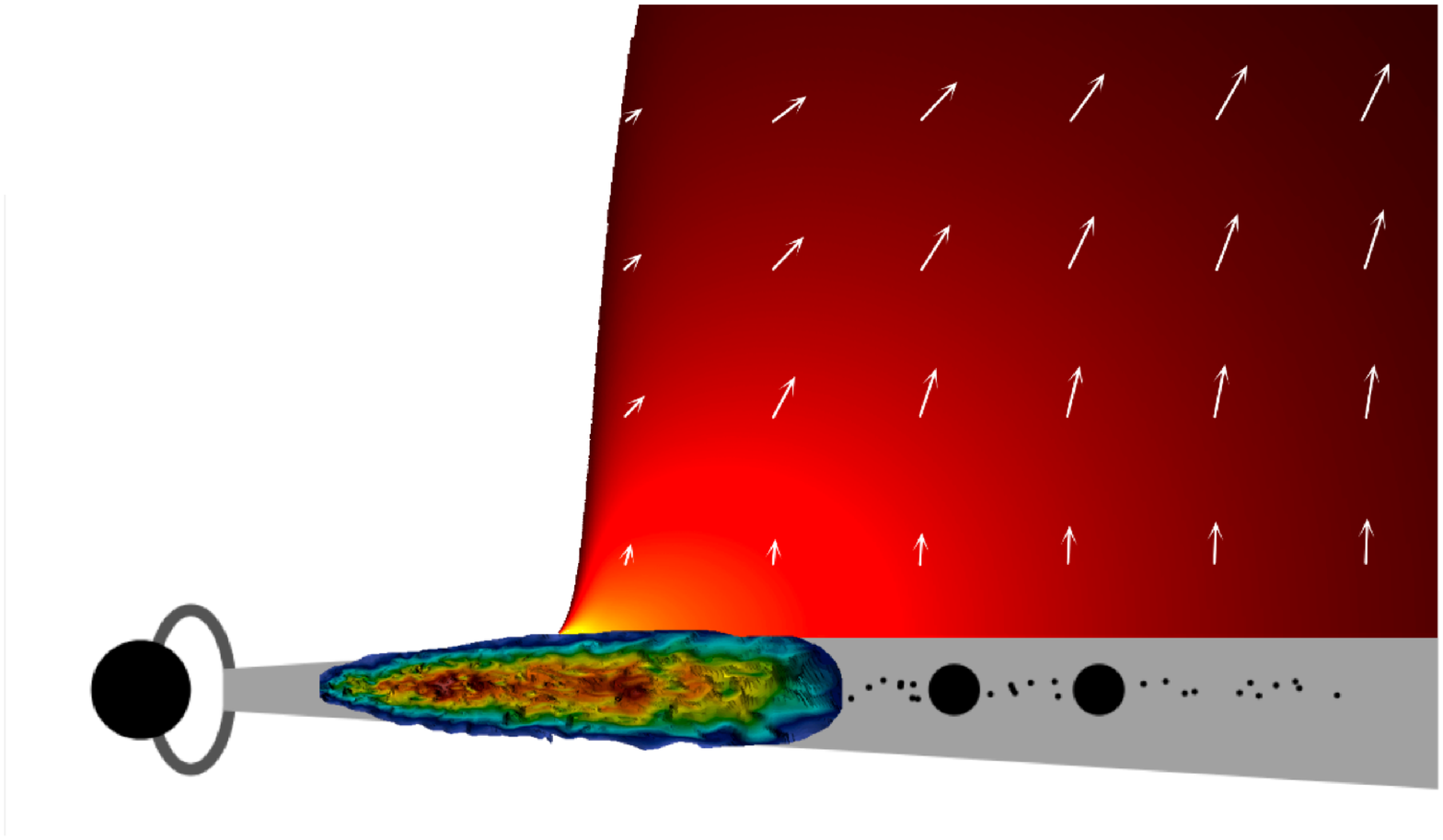} 
 \caption{Schematic picture of an evolving protoplanetary disc.  MHD turbulence drives accretion, while planetesimals and the cores of gas-giant planets form from the agglomeration of dust particles at the disc midplane. UV and X-ray photons from the star irradiate the upper layers of the disc, driving mass loss via a photoevaporative wind. Accretion dominates for most of the disc's lifetime, before photoevaporation takes over at late times and clears the disc.  This simplified picture neglects many details (such as dead zones and MHD-driven winds), but is broadly consistent with observations of T Tauri stars. [Turbulent disc simulation from \citet{beckwith11}; photoevaporative wind simulation from \citet{alexander08b}.] }
   \label{fig:schematic}
\end{center}
\end{figure}


These photoevaporative mass-loss rates, inferred from both theory and observations, are comparable to the observed accretion rates for many T Tauri stars, and are clearly large enough to play a role in the later stages of disc evolution.  However, in order to understand how protoplanetary discs evolve we must consider accretion and photoevaporation simultaneously.  This was first done by \citet{clarke01}, who coupled EUV photoevaporation to a 1-D viscous accretion model.  Models of this type have subsequently been refined and improved by a number of different groups \citep[e.g.,][]{alexander06b,gorti09,owen10,bae13}, and although the details depend on the choice of model parameters and input physics, all photoevaporation/viscous evolution models display the same characteristic behaviour.  Initially the accretion rate is much higher than the photoevaporation rate and the disc evolves viscously, accreting mass inwards and spreading at large radii.  However, as the disc mass is finite the disc accretion rate declines with time, and after a few Myr the accretion rate becomes comparable to the rate of photoevaporative mass-loss.  At this point the wind opens a gap in the disc at $\simeq$\,$R_{\mathrm c}$, and once photoevaporation overcomes accretion the inner disc (inside $R_{\mathrm c}$) is cut-off from re-supply and drains on to the star on its (short) viscous time-scale ($\sim$\,$10^5$yr).  Once the inner disc has been drained photoevaporation rapidly clears the outer disc, and the entire disc is dispersed on a time-scale of a few\,$\times$\,$10^5$yr.  This three-stage evolution scenario (illustrated in Fig.\,\ref{fig:surf_den}) is consistent with disc observations across a wide range in wavelength \citep[e.g.,][]{alexander06b}, and successfully reproduces the two-time-scale behaviour demanded by observations.  We note also that this type of inside-out clearing inevitably gives rise to a short ``inner hole'' phase, and represents a possible mechanism for producing some of the observed transitional discs \citep[e.g.,][]{cieza08}.

Although different photoevaporation models result in qualitatively similar behaviour, the different mass-loss rates and profiles do give rise to significant quantitative differences in how discs evolve.  In particular, the large wind rates predicted for X-ray- and FUV-dominated photoevaporation imply that mass-loss is significant even at early times, and that the total mass ``lost'' to photoevaporation  over the disc's lifetime may be larger than that accreted on to the star.  These models also result in distinctly different properties during the clearing phase (most notably higher disc masses and accretion rates than in the EUV case), and therefore make very different predictions for the properties of transition discs.  However, the dominant uncertainties in this picture of disc evolution remain our lack of understanding of angular momentum transport, and also of protoplanetary disc initial conditions (in particular the initial angular momentum distribution).  

Despite these uncertainties, this evolutionary scenario is now supported by mature theoretical models and a wealth of observational evidence.  For much of the protoplanetary disc lifetime the evolution of the gas disc is essentially governed by the competition between accretion and photoevaporation.  During this time solid material in the disc is agglomerating into progressively larger bodies, and planets are forming and migrating through the disc; this scenario is illustrated schematically in Fig.\,\ref{fig:schematic}.  The processes driving disc evolution and dispersal essentially compete with those of planet formation, depleting the disc and ultimately terminating the epoch of (giant) planet formation.


\section{Implications for planetary systems}
\noindent Having reviewed the physics of protoplanetary disc evolution, I now consider the implications of these results for the formation and evolution of planetary systems.  Disc evolution and dispersal influences planet formation in a number of different ways, by altering the disc's chemistry, changing the dust-to-gas ratio, and ultimately by starving forming planets of the gas reservoir from which they accrete \citep[e.g.,][]{shu93,tb05,gh06}.  Disc evolution also influences the architectures of young planetary systems: the migration of low-mass planets is very sensitive to the local disc structure, and disc clearing halts gas-driven planet migration.  These issues are discussed in detail in \citet{rda_ppvi}; for reasons of length, I focus here on the effects of disc clearing on giant planet migration.


\subsection{Halting planet migration}
\begin{figure}[t]
\begin{center}
 \includegraphics[width=3.4in,angle=270]{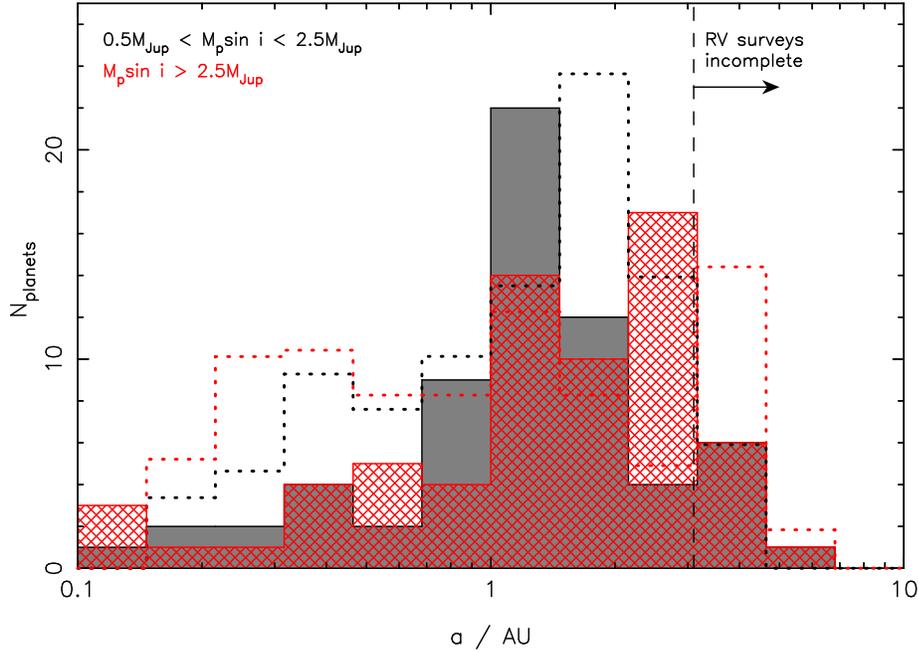} 
 \caption{Exoplanet semi-major axes: the shaded histograms show current exoplanet data \citep[taken from {\tt exoplanets.org};][]{wright11}; the dashed histograms show model predictions from \citet{ap12}. The ``pile-up'' of $\sim$Jupiter-mass planets is plausibly explained by the interaction between migrating planets and gap-opening due to photoevaporation. [Figure adapted from \citet{ap12}.] }
   \label{fig:pileups}
\end{center}
\end{figure}


\noindent It was recognised more than 30 years ago that planets in discs migrate \citep{gt79,gt80}, and when large numbers of exoplanets were discovered in relatively short-period orbits ($P$\,$\ll$\,1yr), migration offered a natural explanation.  However, the predicted migration time-scales are invariably much shorter than observed disc lifetimes \citep[see, e.g.,][]{kley12}, so planet migration must be slowed or stopped in order to produce planets at their observed locations.  Type I migration, which applies to low-mass planets ($\lesssim$\,0.1\Mjup), can be halted or even reversed by changes in the local disc structure \citep{pp09}, but Type II migration, which applies to giant planets ($\gtrsim$\,0.5\Mjup) is driven by the viscous accretion flow and continues as long as the disc accretes.  Disc dispersal offers a obvious means of stopping Type II migration, but the manner and time-scale of disc clearing can have a strong influence on the dynamics of migrating planets.

\citet{armitage02} showed that the distribution of then-known exoplanets was consistent with a simple model, where giant planets form outside the snow-line (at 5--10AU) and undergo Type II migration before being stranded when the disc is dispersed.  Subsequent population synthesis calculations have considered the processes of planet formation in much more detail \citep[e.g.,][]{il04,mordasini09}, but have generally adopted highly simplified models of disc dispersal (such as exponential depletion at all radii).  \citet{aa09} considered the migration of giant planets in a viscous/photoevaporation disc model, and ran large numbers of models to generate statistical samples.  They tested their model against observations of both discs and exoplanets, and showed that this simple scenario successfully reproduces the observed properties of protoplanetary discs and the distribution of giant planets at $\sim$AU radii.  However, as disc dispersal by photoevaporation is not scale-free, it has the potential to leave a signature on the resulting planet distribution.  \citet{ap12} revisited this issue, and found that photoevaporative gap-opening alters planet migration in a manner that is rather sensitive to both the planet mass and the efficiency of accretion across the planet's orbit.  The migration rate of planets close to $R_{\mathrm c}$ changes when the gap opens, and this results in mass-dependent ``deserts'' and ``pile-ups'' in the distribution of planet semi-major axes (see Fig.\,\ref{fig:pileups}).  \citet{ap12} tentatively associated the observed pile-up of $\sim$Jupiter-mass planets at $\sim$1AU with this effect, and suggested that we may be able to link features in the observed exoplanet distribution to the physics of protoplanetary disc evolution.  Other mechanisms, such as planet traps \citep[e.g.,][]{hp12}, can also lead to similar planet pile-ups, and it remains to be seen whether such features in the initial planet distribution are preserved over Gyr time-scales.  However, these results suggest that it may be possible to use the observed properties of exoplanets to inform our understanding of both disc evolution and planetary accretion.


\section{Summary}
\noindent In summary, we now understand that planets form in evolving protoplanetary discs, and that disc evolution plays a major role in both the formation of planets and the subsequent dynamical evolution of planetary systems.  The main processes driving disc evolution on Myr time-scales are accretion, driven by MHD turbuence, and photoevaporation by high-energy radiation from the central star (though magnetically-driven winds may also play a role).  Both accretion and photoevaporation are now observed directly, and there is good agreement between these observations and theoretical predictions. However, while our picture of protoplanetary disc evolution and dispersal is now relatively mature, its implications for planet formation have yet to be explored in detail.  Disc evolution alters the conditions under which planets form, while disc clearing halts planet migration and may leave footprints in the observed distribution of exoplanets.  Future observations and models of both discs and planets will explore these issues further, and allow us to build up a comprehensive picture of the early evolution of planetary systems.


\vspace*{24pt}
\noindent {\bf Acknowledgements} I am grateful to many colleagues and collaborators, especially Ilaria Pascucci, Sean Andrews, Phil Armitage \& Lucas Cieza, for a number of useful and interesting discussions.  I also thank Ilaria Pascucci and James Owen for providing some of the data used in Fig.\,\ref{fig:neii}, and Phil Armitage for the turbulent disc image used in Fig.\,\ref{fig:schematic}. My research is supported by the Science \& Technology Facilities Council (STFC), through an Advanced Fellowship (ST/G00711X/1) and a Consolidated Grant (ST/K001000/1).



\end{document}